\documentclass[aps,prl,showpacs,twocolumn,amsmath,amssymb,superscriptaddress]{revtex4-2}

\usepackage{tabularx}
\usepackage{bm}
\usepackage{graphicx}
\usepackage[scr=esstix]{mathalfa}

\usepackage{hyperref}
\hypersetup{colorlinks=true,urlcolor= blue,citecolor=blue,linkcolor= blue,bookmarksopen=false}

\usepackage{color}

\usepackage{amsmath,mathtools}
\usepackage{multirow}
\usepackage{dcolumn}
\usepackage{amssymb,amscd,xypic,bm,wasysym}
\usepackage{float}
\usepackage{cleveref}
\usepackage[caption=false,position=top,captionskip=0pt,farskip=0pt]{subfig}
\DeclareMathOperator{\Tr}{Tr}

\usepackage{soul}

\begin{document}
	
\title{Hall mass and transverse Noether spin currents in noncollinear antiferromagnets}

\author{Luke Wernert}
\affiliation{Department of Physics, Colorado State University, Fort Collins, CO 80523, USA}
\author{Basti\'an Pradenas}
\affiliation{William H. Miller III Department of Physics and Astronomy, Johns Hopkins University, Baltimore, MD 21218, USA}
\author{Oleg Tchernyshyov}
\affiliation{William H. Miller III Department of Physics and Astronomy, Johns Hopkins University, Baltimore, MD 21218, USA}
\author{Hua Chen}
\affiliation{Department of Physics, Colorado State University, Fort Collins, CO 80523, USA}
\affiliation{School of Materials Science and Engineering, Colorado State University, Fort Collins, CO 80523, USA}

\begin{abstract}
Noncollinear antiferromagnets (AFMs) have recently attracted attention in the emerging field of antiferromagnetic spintronics because of their various interesting properties. Due to the noncollinear magnetic order, the localized electron spins on different magnetic sublattices are not conserved even when spin-orbit coupling is neglected, making it difficult to understand the transport of spin angular momentum. Here we study the conserved Noether current due to spin-rotation symmetry of the local spins in noncollinear AFMs. Interestingly, we find that a Hall component of the spin current can be generically created by a longitudinal driving force associated with a propagating spin wave, inherently distinguishing noncollinear AFMs from collinear ones. We coin the corresponding Hall coefficient, an isotropic rank-4 tensor, as the Hall (inverse) mass, which generally exists in noncollinear AFMs and their polycrystals. The resulting Hall spin current can be realized by spin pumping in a ferromagnet (FM)-noncollinear AFM bilayer structure as we demonstrate numerically, for which we also give the criteria of ideal boundary conditions. Our results shed light on the potential of noncollinear AFMs in manipulating the polarization and flow of spin currents in general spintronic devices.
\end{abstract}

\maketitle
Antiferromagnets (AFMs) with noncollinear magnetic order have recently become a topic of interest in spintronics. In spite of the usual challenge associated with vanishing net magnetization pertinent to all AFMs, which has been significantly mitigated in recent years due to the rapid development of antiferromagnetic spintronics \cite{baltz_2018,manchon_2019,Jungwirth_2018,smejkal_2022,fukami_2020,Han_2023}, the complex magnetic ordering of noncollinear AFMs leads to exotic transport phenomena that open up opportunities otherwise unavailable in common collinear AFMs. A prominent example is the anomalous Hall effect (AHE) in the noncollinear AFM family Mn$_3X$ (where $X$ = Ir, Sn, Ge, etc.)\cite{Hua2014,Mn3GeTHEORY,Liu,JpnMn3Sn,Mn3GeEXP,Mn3IrEXP,Mn3PtEXP,smejkal_2022} as well as other transport and optical properties with the same symmetry requirements as the AHE \cite{JpnMn3Sn,Mn3GeEXP,Mn3IrEXP,Mn3PtEXP,Nernst2017,chen_2020,Kerr2015,Kerr2018}. The low symmetry of the magnetic structure also allows the existence of the magnetic spin-Hall (MSHE) and inverse spin-Hall (MISHE) effects \cite{Kimata2019,Kondou_2021, SHE2017, SHE2018}, anisotropic magnetorestriction and piezomagnetic effects \cite{Ikhlas_2022,Zuniga-Cespedes_2023}, and nontrivial spin-transfer torques \cite{SpinPolCurr,ghosh_2022}, etc. Some of the above properties are investigated using low-energy Hamiltonian near the Weyl nodes of Mn$_3X$ \cite{Weyl2014,Weyl2017,Liu,Kuroda2017}. In addition to Mn$_3X$, transport phenomena in other noncollinear AFMs such as antiperovskite Mn$_3AB$ ($A =$ Ga, Ni, Cu; $B = $ C, N) \cite{AntiperovAHE1,AntiperovAHE2,AntiperovAHE3}, orthoferrites \cite{Li2019}, vector spin Seebeck effect \cite{xu_2022}, and many effective Ising magnets \cite{Fujita2015,Wang2020,Xiao_2020,zhao_2020,Zhao2024} have also garnered significant interest. 

In a magnet, spin currents can be carried by itinerant electrons or magnons. Both mechanisms have been well characterized for ferromagnets (FMs) and collinear AFMs with isotropic (exchange) interactions \cite{DefofSpinCurr}. The magnetic order parameter in the form of uniform or staggered magnetization lowers the SO(3) symmetry of global spin rotations to that about the magnetization direction. Consequently, only the longitudinal component of spin current is usually considered (an exception is spin superfluids in easy-plane magnets where it is carried by the magnetic ground state). Even with finite spin-orbit coupling, an approximate diffusive picture of spin transport can usually be established, historically playing a powerful role in discovering and understanding many remarkable phenomena in FM- and AFM-based spintronics \cite{STT1996, STTexp, SpinHallEffect, tserkovnyak_2005, EarlyAFMCurrPaper, Spincurrenttrans, Han_2023}. In contrast, in noncollinear AFMs, the magnetic order parameter breaks spin rotation symmetry completely, making it unclear whether a conserved spin current can even be defined. 

\begin{figure}[ht]
	\centering
	\includegraphics[width=0.7\linewidth]{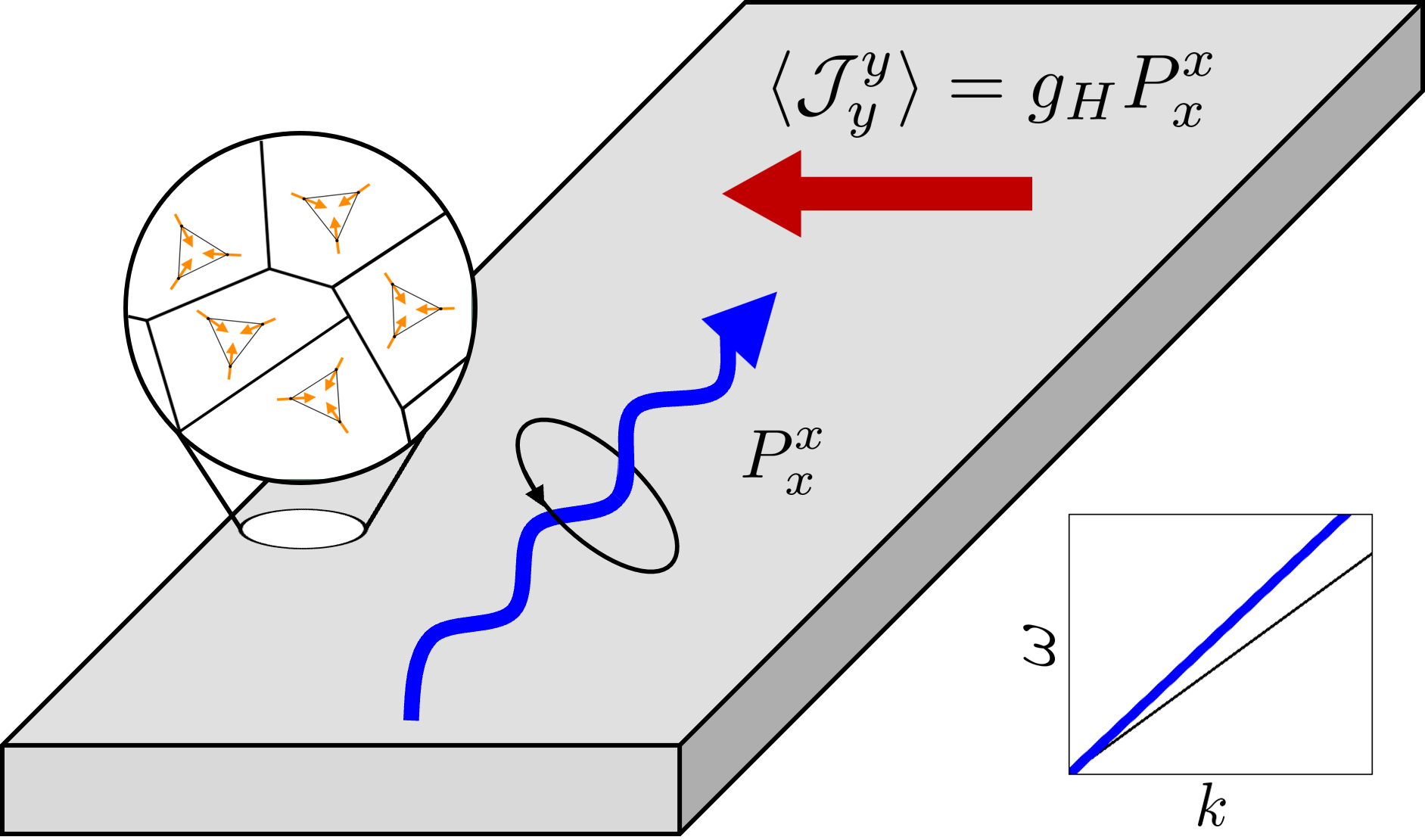}
	\caption{Hall spin current generated by the Hall mass $g_{\rm H}$ of a polycrystalline noncollinear AFM. The blue wavy line represents circularly polarized spin waves (thick blue line in the magnon dispersion curves in the inset) propagating along $x$. The red arrow corresponds to the d.c. dynamical Hall spin current flowing along $y$. The Hall mass $g_{\rm H}$ also leads to the splitting between the transverse and longitudinal (solid black line in inset) polarization magnon modes. }
	\label{Fig:schematic}
\end{figure}

Despite this conceptual difficulty, recent discoveries in noncollinear AFMs, including MSHE/MISHE \cite{Kimata2019, Kondou_2021}, and tunneling magnetoresistance \cite{OctupoleMagRes,RoomTempMagRes} have been interpreted using a spin current language. These heuristic arguments suggest that some form of a conserved spin current might nonetheless exist in noncollinear AFMs. In this context, the difference between the noncollinear AFMs and collinear magnets mentioned above raises the question of whether there are any spin current related properties that fundamentally distinguish the two. Here, we provide an answer to this question: In FMs and collinear AFMs, the spin current flows along the spatial gradient of the magnetic order parameter, while this is generally not the case in noncollinear AFMs. As illustrated in Fig.~\ref{Fig:schematic}, a spatial gradient of the magnetic order in the form of spin waves propagating in the $x$ spatial direction may induce a spin current flowing along $y$, reminiscent of a Hall effect. 

In this Letter, we first derive a conserved current for global spin rotations by applying Noether's theorem to a continuum theory \cite{Andreev1980,hydrodyn,halperin_1977,Dombre, SpinCurBook} of a kagome Heisenberg antiferromagnet exemplified by the Mn$_3X$ family \cite{Ulloa, Gom2012, Gom2015, Rodrigues_2022, DMIandotherIntsMn3XAB, nikolic_2020, li_kovalev_2021, lund_2021, Pradenas:unpub}. We then show that the Noether spin current can have a component transverse to the gradient of the order parameter, i.e. a Hall spin current. To verify its existence, we have simulated the spin dynamics in a lattice model consisting of an interface between an FM driven by external a.c. fields and a noncollinear kagome AFM, guided by analytic insights on the ideal boundary conditions for magnon transmission. The obtained d.c. spin current on the AFM side due to the injected spin waves from the FM side indeed has a component flowing parallel to the interface. The Hall spin currents originate from off-diagonal components of the inverse effective mass tensor in the continuum Lagrangian. We name in particular the isotropic part of such off-diagonal components the Hall (inverse) mass, which resembles the charge Hall vector and the scalar spin Hall conductivity in their persistence even in the absence of crystallinity. Finally we discuss possible experimental approaches that can detect the Hall mass.

\begin{figure}[ht]
	\centering
	\subfloat[]{\includegraphics[width=0.52\linewidth]{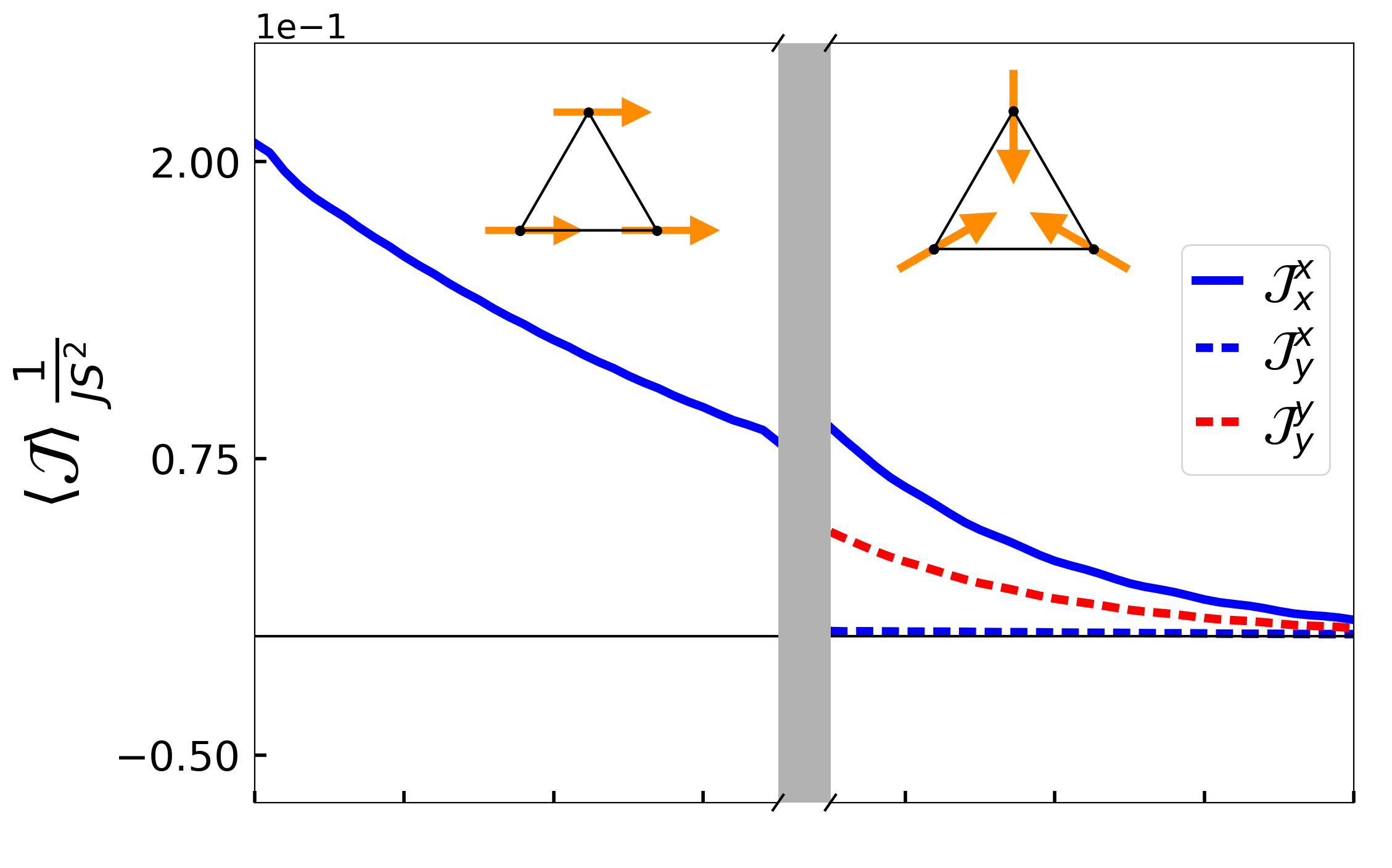}} 
        \subfloat[]{\includegraphics[width=0.48\linewidth]{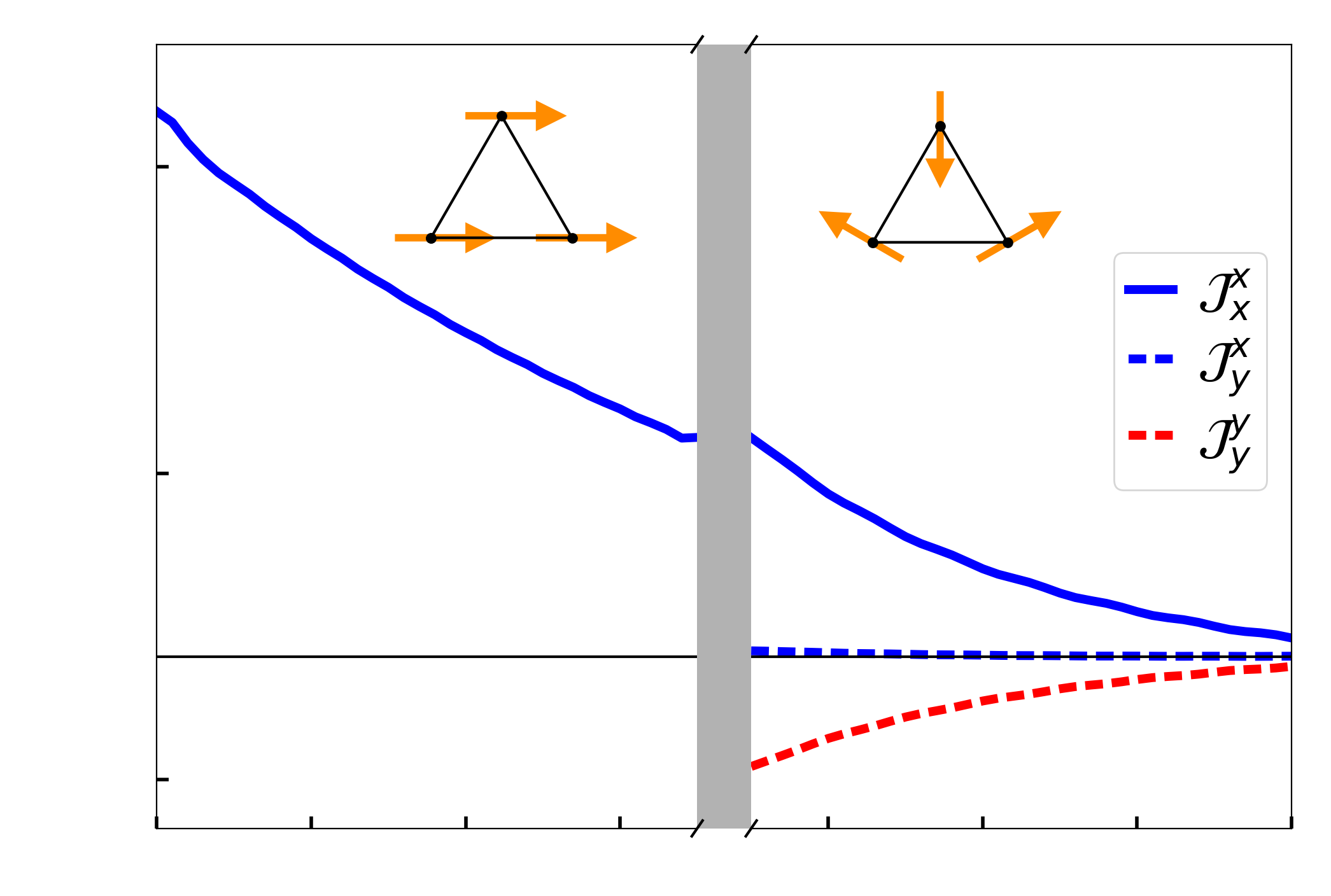}} \\
 	\subfloat[]{\includegraphics[width=0.51\linewidth]{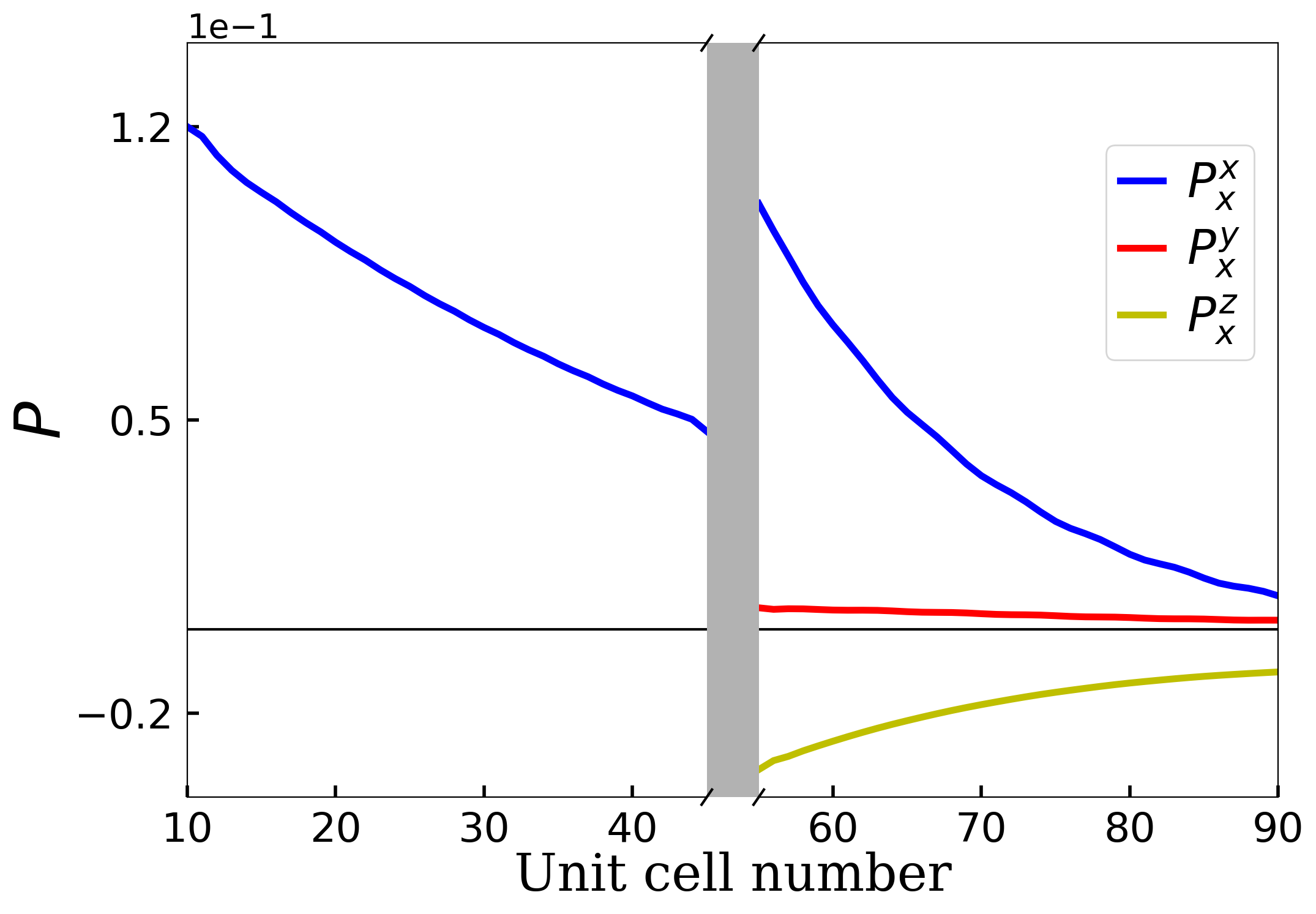}}
	\subfloat[]{\includegraphics[width=0.49\linewidth]{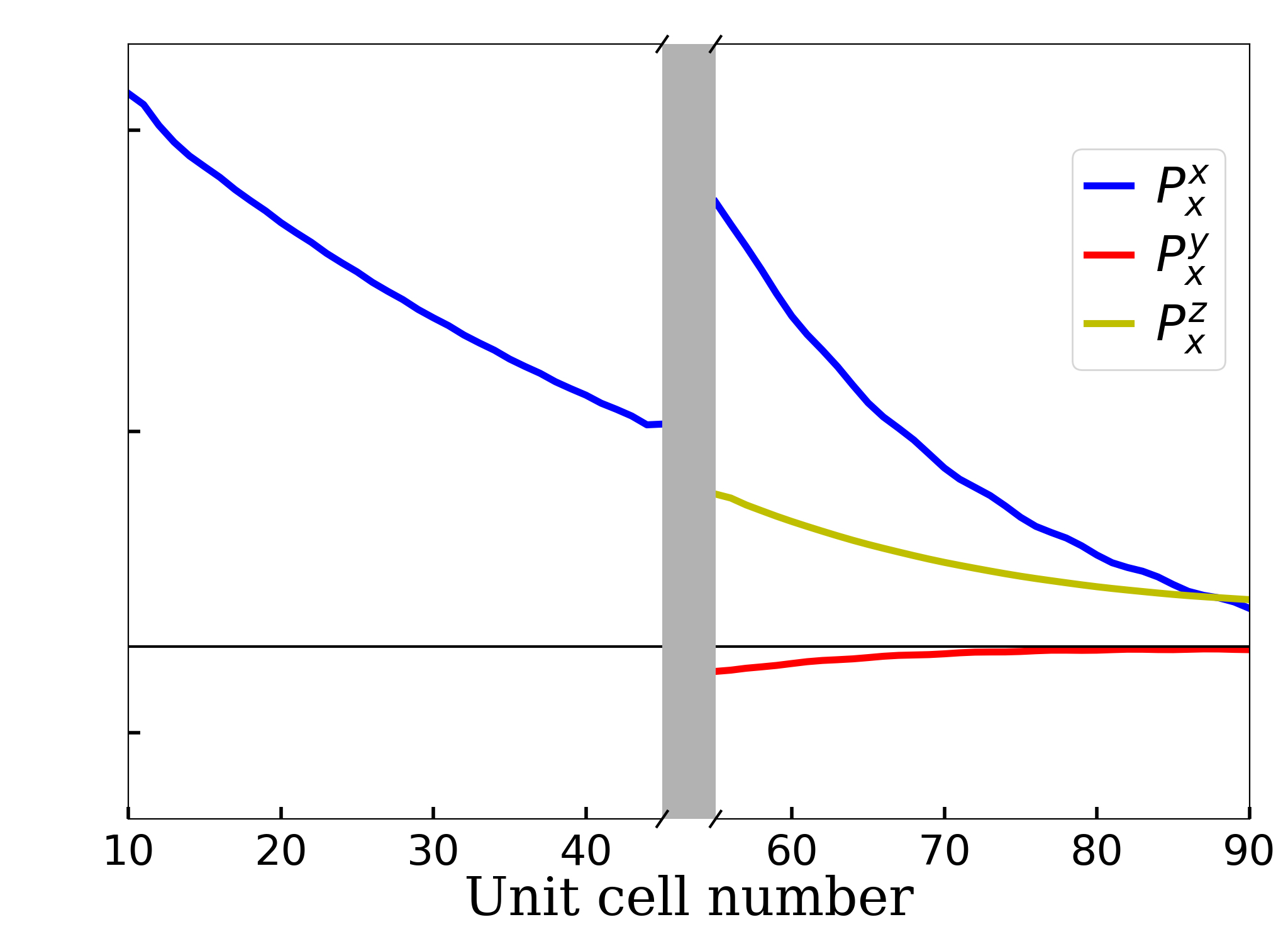}}
	\caption{(a) and (b), Dynamical d.c. spin currents in (a) FM-Mn$_3$Ir and  (b) FM-Mn$_3$Sn interfaces outside of the interface region (gray rectangles). (c) and (d), Spin current driving force $P$ for (c) FM-Mn$_3$Ir and (d) FM-Mn$_3$Sn interfaces.}
	\label{Fig:Iruni}
\end{figure}

\emph{Noether spin current of noncollinear AFMs.---}We first use a prototypical two-dimensional model \cite{Dombre,Ulloa} to study some generic properties of Noether spin currents in noncollinear AFMs. The model has classical spins of length $S$ with antiferromagnetic nearest-neighbor exchange coupling $J$ on a kagome lattice. We consider two representative noncollinear AFM ground states, direct [Fig.~\ref{Fig:Iruni} (a) inset] and inverse triangular order [Fig.~\ref{Fig:Iruni} (b) inset], relevant to that in Mn$_3X$. 

The low-energy behavior of the model is captured by a continuum Lagrangian obtained through gradient expansion \cite{Dombre,Ulloa,li_kovalev_2021,lund_2021,Rodrigues_2022,Pradenas:unpub,supp}. The magnetization fields of the three sublattices are coplanar and point at angles of $120^\circ$ to each other, forming a rigid body. The orientation of this object can be encoded in terms of a spin frame---three mutually orthogonal unit vectors $\mathbf n_\alpha$, where $\alpha = x, y, z$, rigidly attached to the magnetic order parameter \cite{Pradenas:unpub,supp}. The low-energy Lagrangian is
\begin{eqnarray}\label{eq:Lsf}
\mathcal{L} = \frac{\rho}{4} \partial_t \mathbf n_\alpha \cdot \partial_t \mathbf n_\alpha - \frac{1}{4} \Gamma_{ab}^{\alpha\beta} \partial_a\mathbf n_\alpha \cdot \partial_b \mathbf n_\beta
\end{eqnarray}
Here the inertia density $\rho = \frac{1}{2JA_c}$ is related to the paramagnetic susceptibility. $A_c = 2\sqrt{3}a_0^2$ is the unit cell area and $a_0$ is the nearest-neighbor distance. Latin indices $a$ and $b= x, y$ label spatial directions and Greek indices denote spin components. Summation over doubly repeated indices is implied. The spin-frame vectors can be chosen as certain superpositions of sublattice magnetizations in such a way that $\mathbf n_x$ and $\mathbf n_y$ transform under point-group symmetries of the lattice in the same way as spatial gradients $\partial_x$ and $\partial_y$ (hence the labels) \cite{Pradenas:unpub}. The third spin-frame vector $\mathbf n_z = \mathbf n_x \times \mathbf n_y$ then has the meaning of the vector spin chirality. The fourth-rank tensor $\Gamma$ is generally symmetric under the simultaneous exchange of the Latin and Greek indices \cite{supp}. For the kagome models considered above we found,
\begin{equation}\label{eq:Gamma2D}
\Gamma_{ab}^{\alpha\beta}
= \eta\frac{\sqrt{3}}{4}JS^2 (\delta_{a\alpha} \delta_{b\beta}
+ \delta_{a\beta} \delta_{b\alpha}),
\end{equation}
where $\alpha$ and $\beta$ take on values $x$ or $y$; it vanishes if either $\alpha = z$ or $\beta = z$. $\eta=\pm 1$ depends on the spin order: For the direct triangular order $\eta = 1$, while for the inverse triangular order $\eta = 1 (-1)$ when $\alpha = \beta (\alpha\neq \beta)$. Alternatively, the spin-frame vectors $(\mathbf n_\alpha)_m$ can be understood as column vectors of a rotation matrix $R_{m\alpha}$ \cite{supp}, which leads to an equivalent form of Eq.~\eqref{eq:Lsf} \cite{Dombre,Ulloa}:
\begin{equation}\label{Eqn:L}
\mathcal{L} = 
- \frac{\rho}{4} \Tr{[(R^{-1} \partial_t R)^2]}
+\frac{1}{4} \Tr[\Gamma_{ab}(R^{-1} \partial_a R)(R^{-1} \partial_b R)]
\end{equation}
where the trace is over spin indices. The latter will be omitted below for brevity when possible. 

The equations of motion for the spin frame, obtained by minimizing the action with respect to the fields $\mathbf n_\alpha$ while maintaining the orthonormality constraints $\mathbf n_\alpha \cdot \mathbf n_\beta = \delta_{\alpha\beta}$ \cite{Pradenas:unpub}, are
\begin{equation}
\frac{\rho}{2} \mathbf n_\alpha \times \partial_t^2 \mathbf n_\alpha 
- \frac{1}{2} \Gamma_{ab}^{\alpha\beta} \,
\mathbf n_\alpha \times \partial_a \partial_b \mathbf n_\beta = 0.
\label{eq:LL}
\end{equation}

Lagrangian \eqref{eq:Lsf} is invariant under global spin rotations, $\delta \mathbf n_\alpha = \boldsymbol \theta \times \mathbf n_\alpha$ for an infinitesimal rotation angle $\boldsymbol \theta$. By Noether's theorem, the spin current flowing along spatial direction $a$ is
\begin{equation}\label{Eqn:NoetherCurrent}
\boldsymbol{\mathcal{J}}_a = 
\mathbf n_\alpha \times 
\frac{\partial \mathcal L}{\partial(\partial_a \mathbf n_\alpha)}   
= - \frac{1}{2} \Gamma_{ab}^{\alpha\beta} \,
\mathbf n_\alpha \times \partial_b \mathbf n_\beta.
\end{equation}
The spin density is similarly obtained as 
\begin{eqnarray}\label{Eqn:AngMomDen}
\boldsymbol{\mathcal{J}}_0 = 
 \frac{\rho}{2} \mathbf n_\alpha \times \partial_t \mathbf n_\alpha
= \rho \boldsymbol \Omega 
= \frac{1}{A_c} \sum_{i=1}^3 \mathbf S_i,
\end{eqnarray}
where $\boldsymbol \Omega$ is the rotation frequency of the spin frame. The last equality comes from the equation of motion for the canting field \cite{supp}. The continuity equation for the spin current, $\partial_t \boldsymbol{\mathcal{J}}_0 + \partial_a \boldsymbol{\mathcal{J}}_a = 0$, follows directly from the equation of motion \eqref{eq:LL}.

The Noether spin current Eq.~\eqref{Eqn:NoetherCurrent} generally becomes nonzero whenever the spin configuration is nonuniform, even if it is static. For example, consider a static spin configuration with the spin frame twisting about $\mathbf n_x$ as one moves along the spatial $x$ direction, $\partial_x \mathbf n_\alpha = \partial_x \phi \, \mathbf n_x \times \mathbf n_\alpha$, where $\phi(x)$ is a twist angle. The only nonzero spin current components are $\boldsymbol{\mathcal{J}}_y = \pm\frac{\sqrt{3}}{8}JS^2 \, \partial_x \phi \, \mathbf n_y$ whose spatial direction is orthogonal to the gradient of $\phi$, and are therefore reminiscent of a Hall current. This will be our focus below.

\emph{Hall mass and transverse Noether spin currents.---}In this section we illustrate the richness of the Hall spin currents mentioned above by considering dynamical noncollinear spins which are more relevant to typical spintronics experiments. We consider d.c. spin currents due to spin waves, for which $\mathbf{n}_\alpha (\mathbf r, t) = \mathbf n^0_\alpha + \bm \theta(\mathbf r, t) \times \mathbf n^0_\alpha$, $\mathbf n^0_\alpha$ being the ground state spin frame vectors and $\boldsymbol{\theta} (\mathbf{r}, t) = {\rm Re}[\boldsymbol{\theta} e^{i(\mathbf{k}\cdot\mathbf{r}-\omega t)}]$. Taking a time average on both sides of Eq.~\eqref{Eqn:NoetherCurrent} and subtracting any static contributions in equilibrium, we obtain
\begin{eqnarray}\label{Eqn:DynJ}
		\langle \mathcal{J}_a^\alpha \rangle =\Gamma_{ab}^{\gamma\beta} \left[ - \frac{1}{2} \left\langle\mathbf n_\gamma \times \partial_b \mathbf n_\beta \right\rangle_\alpha \right ]\equiv \mathrm{Tr} \left( \Gamma_{ab} \langle \mathcal{P}^\alpha_b \rangle \right)
\end{eqnarray}
which resembles a linear response. Indeed, we show in \cite{supp} that $\mathcal{P}_b^\alpha$ can be understood as an SO(3) gauge potential \cite{Hill_2021} due to spatial translation along $b$ projected onto the $\alpha$-th spin angular momentum component, and can therefore be viewed as a spin current driving force \cite{nikolic_2020}. Moreover, $\langle \mathcal{P} \rangle$ can be compactly expressed using time-averaged densities of energy, linear momentum, and spin for spin waves, whose behavior in scattering events can be understood separately, as presented below.

From Eq.~\eqref{Eqn:AngMomDen}, the time averaged spin density is 
\begin{eqnarray}\label{eq:J0avg}
\langle\mathcal{J}_0^\gamma \rangle= -\frac{\rho\omega}{4}\mathrm{Im}(\boldsymbol{\theta} \times \boldsymbol{\theta}^*)_\gamma
\end{eqnarray}
Note that due to the last equality of Eq.~\eqref{Eqn:AngMomDen}, the time-averaged canting of the noncollinear spins is equal to the $\langle\mathcal{J}_0^\gamma \rangle$ above and represents the other part of the angular momentum that is carried by the spin waves, which will be discussed in a future work. Eq.~\eqref{eq:J0avg} suggests that linearly polarized AFM spin waves, for which $\bm \theta$ are all real and $\boldsymbol{\theta} \times \boldsymbol{\theta}^* = 0$, cannot carry spin. However, the kagome AFM models considered here all have degenerate spin wave modes that can be circularly polarized \cite{Pradenas:unpub}. More general cases will be discussed in the next section. The time-averaged linear momentum ($\mathcal{T}_{b0}$) and energy ($\mathcal{T}_{00}$) densities are \cite{supp}
\begin{eqnarray}\label{eq:T0mu}
	\langle \mathcal{T}_{b0} \rangle 
	= 
\frac{\rho\omega}{2}
	|\boldsymbol{\theta}|^2 k_b,\; \langle \mathcal{T}_{00} \rangle = \frac{\rho\omega^2}{2} |\boldsymbol{\theta}|^2
\end{eqnarray}
where $|\boldsymbol{\theta}|^2 = \boldsymbol{\theta} \cdot \boldsymbol{\theta}^*$. As a result,
\begin{eqnarray}\label{eq:Psw}
	\langle\mathcal{P}^\alpha_b\rangle_{\beta\gamma} =
	\frac{1}{\rho} \frac{\langle\mathcal{J}_0^\gamma \rangle \langle \mathcal{T}_{b0} \rangle }{\langle \mathcal{T}_{00} \rangle} \delta_{\alpha\beta} \equiv P_b^\gamma \delta_{\alpha\beta},
\end{eqnarray}

We therefore have $\langle \mathcal{J}_a^\alpha\rangle = \Gamma_{ab}^{\alpha\gamma}P_b^\gamma$ which even better illustrates the meaning of $\Gamma$ as a response tensor than Eq.~\eqref{Eqn:DynJ}. It is then remarkable that, due to the off-diagonal components of $\Gamma$, neither the spin nor the spatial directions of the spin current have to be aligned with that of the driving force, or equivalently with that of the spin wave's spin and linear momentum, suggesting the existence of Hall spin currents.

To demonstrate the Hall spin currents explicitly, we consider a bilayer system consisting of an FM layer interfaced with a noncollinear AFM layer in 2D, both on a kagome lattice (Fig.~\ref{Fig:Iruni} insets). We then solve the linearized LLG equation for harmonic excitations created by an external a.c. magnetic field applied on a few leftmost layers on the FM side (see the End Matter for details). In the FM layer, the excitations correspond to FM spin waves propagating towards the interface, carrying a spin angular momentum component in the static magnetization direction, set to $\hat{\mathbf x}$. This is clearly shown by the d.c. spin currents on the FM side in Fig.~\ref{Fig:Iruni}. However, the spin currents on the AFM side have several components not naively expected from that on the FM side, and the transversely flowing $\mathcal{J}_y^y$ component has opposite signs for the direct and inverse triangular noncollinear states. 

To understand this nontrivial behavior, we plot in Figs.~\ref{Fig:Iruni} (c) and (d) the corresponding driving force $P$ in both systems. Different from $\langle\mathcal{J}\rangle$, both sides of the interface have the same dominant component of $P$, $P_{x}^x$, as determined by the linear momentum and spin directions of the propagating spin waves created by the a.c. field. ($P_{x}^y$ and $P_x^z$ arise due to uncontrolled scattering at the interface region.) The discontinuity in the size of $P$ across the interface is reasonable since linear momentum of the spin waves is not conserved in interface scattering. 

Comparing Figs.~\ref{Fig:Iruni} (a) and (b) with (c) and (d), one can find that the nonzero components of $\langle\mathcal{J}\rangle$ are straightforwardly obtained from $P_{x}^x$ and $P_x^y$ together with the nontrivial components of $\Gamma$ in Eq.~\eqref{eq:Gamma2D}. ($P_x^z$ does not contribute since $\Gamma$ has no $z$-spin indices.) For example, the $\langle\mathcal{J}_y^y\rangle$ component is due to $\Gamma_{yx}^{yx} =\pm\frac{\sqrt{3}JS^2}{4}$, whose different signs for the two types of order lead to the opposite $\langle\mathcal{J}_y^y\rangle $ in Figs.~\ref{Fig:Iruni} (a) and (b). Conversely, the $\langle\mathcal{J}^x_y\rangle$ component has the same sign in both systems due to ${P}^y_x$ and $\Gamma_{xy}^{yx} = \Gamma_{yx}^{yx}$ both changing sign.

The above observation suggests that it is meaningful to associate certain components in $\Gamma$ with an analogue of the Hall effect for the Noether spin current. Since $\Gamma$ has the units of inverse mass per area or volume (times $\hbar^2$), we name the part of $\Gamma$ responsible for the Hall spin currents the Hall (inverse) mass. However, it is not as simple as deeming the off-diagonal components such as $\Gamma_{xy}^{xy}$ the Hall mass since the individual components change their values under rotations of the coordinate system, while both the charge Hall effect and the spin Hall effect have certain invariance under coordinate transformations. More specifically, the charge Hall effect is a pseudovector $\sigma_{\rm H}^{a} = \frac{1}{2}\epsilon_{abc}\sigma_{bc}$ \cite{chen_2022}, with $\sigma_{bc}$ the conductivity tensor, whose length is invariant under O(3) transformations; the spin Hall effect has an isotropic part $\frac{1}{6}\epsilon_{abc}\sigma_{abc}^{\rm s}$, with $\sigma_{abc}^{\rm s}$ the spin conductivity tensor, which is a scalar and invariant under O(3) transformations as well. In the next section we propose an appropriate definition of the Hall mass based on a general field theory of noncollinear AFMs.

\emph{General definition of Hall mass in noncollinear AFMs.---}We first generalize our discussion to arbitrary single-crystalline noncollinear AFMs in three dimensions. For concreteness we assume the local spins have length $S$ and pairwise Heisenberg exchange couplings. The $\Gamma$ tensor of such a noncollinear AFM can be obtained through gradient expansion as \cite{supp}
\begin{eqnarray}\label{eqn:gammaorig}
	\Gamma_{ab}^{\alpha\beta} = -\frac{S^2}{V_c}\sum_{j,pq} J_{0p,jq} (\mathbf r_{0p,jq})_a (\mathbf r_{0p,jq})_b {m}_p^\alpha {m}_q^\beta
\end{eqnarray}
where $V_c$ is the volume of the unit cell, $i,j$ label unit cells, $p,q$ label sublattices, $\hat{\mathbf m}_p$ is a unit vector along the spin direction on sublattice $p$, and $\mathbf r_{ip,jq}$ is the position vector of site $jq$ relative to site $ip$. Such a $\Gamma$ is symmetric under separate permutations of its spatial and spin indices and is also independent of unit cell choices \cite{supp}. 

The $\Gamma$ as defined in Eq.~\eqref{eqn:gammaorig} generally has off-diagonal components whose values will also change under rotations of the coordinate system. Moreover, the low-energy spin wave modes that depend on $\Gamma$ are in general non-degenerate and linearly polarized, different from that in ferromagnets, making them unable to carry spin according to Eq.~\eqref{eq:J0avg}. (Spin can also be carried by large-angle precession, or supercurrents \cite{sonin_2010,konig_2001,Nogueira_2004,so_2014,chen_2014_2,so_2014_2,Ochoa_2018,li_kovalev_2021}, of the noncollinear spins, which we do not discuss in this work.) Both of these hurdles can be overcome if we consider a polycrystal of the given noncollinear AFM. When the grain sizes of the polycrystal are smaller than the typical wavelengths of the spin wave modes, its low-energy Lagrangian should have the same form as Eq.~\eqref{Eqn:L}, but with $\rho$ replaced by an effective isotropic paramagnetic susceptibility $\bar{\rho}$ and $\Gamma$ by an angular-averaged $\bar{\Gamma}$ that depends only on two parameters \cite{power_2003,supp}:
\begin{eqnarray}\label{eq:Gammabar}
    \bar{\Gamma}_{ab}^{\alpha\beta} = g_{\rm H} (\delta_{a\alpha} \delta_{b\beta}+ \delta_{a\beta}\delta_{b\alpha}) + g_0 \delta_{ab}\delta_{\alpha\beta}
\end{eqnarray}

We define the $g_0$ and $g_{\rm H}$ in Eq.~\eqref{eq:Gammabar} as the longitudinal and the Hall mass, respectively, since they are now invariant under rotations. (Note that $\delta_{ab}\delta_{\alpha\beta}$, $\delta_{a\alpha}\delta_{b\beta}$, and $\delta_{a\beta}\delta_{b \alpha }$ are the only three linearly independent isotropic rank-4 tensors in three dimensions.) In terms of the $\Gamma$ of the corresponding single crystal, $g_{\rm H} = \frac{1}{10} \left(\Gamma^{ab}_{ab} - \frac{1}{3}\Gamma^{aa}_{bb} \right)$, $g_0 = \frac{2}{15} \left(\Gamma^{aa}_{bb} -\frac{1}{2}\Gamma^{ab}_{ab}\right)$. 

The physical meaning of the Hall mass $g_{\rm H}$ can be understood from the spin waves of isotropic noncollinear AFMs. Using the isotropic $\bar{\Gamma}$ and the polycrystal Lagrangian, one can easily get the spin wave dispersion $\omega_i = c_i k$, where $i={\rm I, II, III}$ and
\begin{eqnarray}\label{eq:eigisotropic}
    c_{\rm I} = \sqrt{g_0/\bar{\rho}},\quad
    c_{\rm II, III}= \sqrt{(g_0+g_{\rm H})/\bar{\rho}}
\end{eqnarray}
The mode I has its polarization parallel to $\mathbf k$ while II, III have transverse polarizations, similar to phonons in isotropic elastic media \cite{supp}. The two transverse modes can thus be circularly polarized and carry the spin that is parallel to the propagation direction. Similar to acoustic phonons, the transversely polarized modes do not have to have the same velocity as the longitudinally polarized mode. The difference in their velocities is essentially set by the Hall mass $g_{\rm H}$ (Fig.~\ref{Fig:schematic}). 

Since $g_{\rm H}$ is an isotropic quantity, its existence cannot be precluded by any magnetic space group symmetry. Therefore $g_{\rm H}$ is generally nonzero in any noncollinear AFM, similar to the situation of SHE generally existing in any, even amorphous, nonmagnetic conductors. The sign of $g_{\rm H}$ therefore serves as a criterion to classify all noncollinear AFMs. 

\emph{Discussion.---}To detect the Hall mass, one can intentionally excite circularly polarized spin waves from the superposition of the two transversely polarized modes, which carry spins parallel to the propagation direction. $g_{\rm H}$ then leads to Hall spin currents that also have the spin direction aligned with the current flow, i.e., ``radial'' spin currents such as the $\langle \mathcal{J}_y^y\rangle$ in Fig.~\ref{Fig:Iruni}. Such spin currents may be detected by ISHE in a low-symmetry crystal \cite{Taniguchi_2015,Humphries2017,SpinPolCurr,MacNeill2017,Kimata2019}. Alternatively, if the noncollinear AFM under consideration has spin wave modes that can carry $y$-spin, one can pump $y$-spin from the FM side to create $\langle \mathcal{J}_{y}^x\rangle$ which is then detectable by ISHE, since $\Gamma_{yx}^{yx} = \Gamma_{yx}^{xy}$. Such a phenomenon is more analogous to the spin swapping effect \cite{lifshits_2009,Lin_2022} despite the different microscopic origins. Alternatively, motivated by Eq.~\eqref{eq:eigisotropic}, one can use spectroscopic methods, such as inelastic magnetic neutron or x-ray scattering, to probe the difference between the velocities of the low-energy modes of powder samples of noncollinear AFMs.

In real 3D noncollinear AFMs the $\Gamma$ tensor can be quite different from those of our 2D kagome models. One can use Eq.~\eqref{eqn:gammaorig} to calculate $\Gamma$ for a given system once the spin Hamiltonian is determined from, e.g., inelastic neutron diffraction and linear spin-wave theory. For example, we have used Eq.~\eqref{eqn:gammaorig} to calculate $\Gamma$ for cubic and hexagonal Mn$_3X$ by keeping the nearest-neighbor exchange coupling only \cite{supp}. We found that even when the individual kagome planes in these materials have the same orientations as that in our 2D models, the 3D materials have different $\Gamma$ components. Nonetheless, the size of the off-diagonal components is usually of the same order as the diagonal components, suggesting that the Hall spin current carried by circularly polarized spin waves should be comparable to the longitudinal spin current. 

The developed formalism of conserved spin current carried by coherent dynamics of noncollinear AFMs is useful beyond the discussion of transverse currents. It can be combined with similar theories of exchange spin currents \cite{SpinCurBook} for collinear magnets together with appropriate boundary conditions to quantitatively describe spin angular momentum transport in a variety of multi-layer systems with complex magnetic ordering. 


\begin{acknowledgements} 
The authors are grateful to Collin Broholm and Satoru Nakatsuji for useful discussions. L.W. and H.C. acknowledge support by NSF CAREER grant DMR-1945023. B.P. and O.T. acknowledge support by the U.S. Department of Energy under Award No. DE-SC0019331 and by the U. S. National Science Foundation under Grants No. PHY-1748958 and PHY-2309135. This work was performed in part at Aspen Center for Physics, which is supported by National Science Foundation grant PHY-2210452.
\end{acknowledgements}

\section*{End Matter: Spin wave transmission}

In this End Matter we provide details on the numerical calculations of d.c dynamical spin currents in an FM/noncollinear AFM interface as well as qualitative criteria for optimizing spin wave transmission through such interfaces. The bilayer structure is modeled by a $100\times 1$ kagome lattice strip, created using the rectangular unit cell spanned by 
\begin{eqnarray}
	\mathbf a_1 = a\hat{x},~\mathbf a_2 = \sqrt{3}a\hat{y}
\end{eqnarray}
with lattice constant $a=1$. Uniform FM moments along $\hat{x}$ are assigned to the left half of the strip, and uniform noncollinear AFM states (either the direct or the inverse triangular order) are assigned to the right half, with periodic boundary conditions along $\hat{y}$. Sites on the FM (AFM) side were given FM (AFM) nearest neighbor coupling, with $J_{\rm FM} = -1$ and $J_{\rm AFM}=1$, except for the sites immediately at the interface which we detail below. No anisotropies are considered for the FM part, while $K = 0.01$, and $K_z = 0.00125$, strengths of easy-axis and easy-plane anisotropies as defined in \cite{supp}, are used for the AFM side. 

The efficiency of spin wave transmission across an FM-noncollinear AFM interface critically depends on the boundary condition. Such problems can generally be formulated as quasi-1D scattering, with the boundary condition derived from coarse-graining or based on symmetry \cite{Spincurrenttrans, supp}. Using this approach, we have shown \cite{supp} that the ideal boundary condition relevant to FM/noncollinear AFM interfaces is when the latter has a fully compensated surface so that its order parameter is not pinned by the FM magnetization. 

\begin{figure}[ht]
	\centering
	\includegraphics[width=0.6\linewidth]{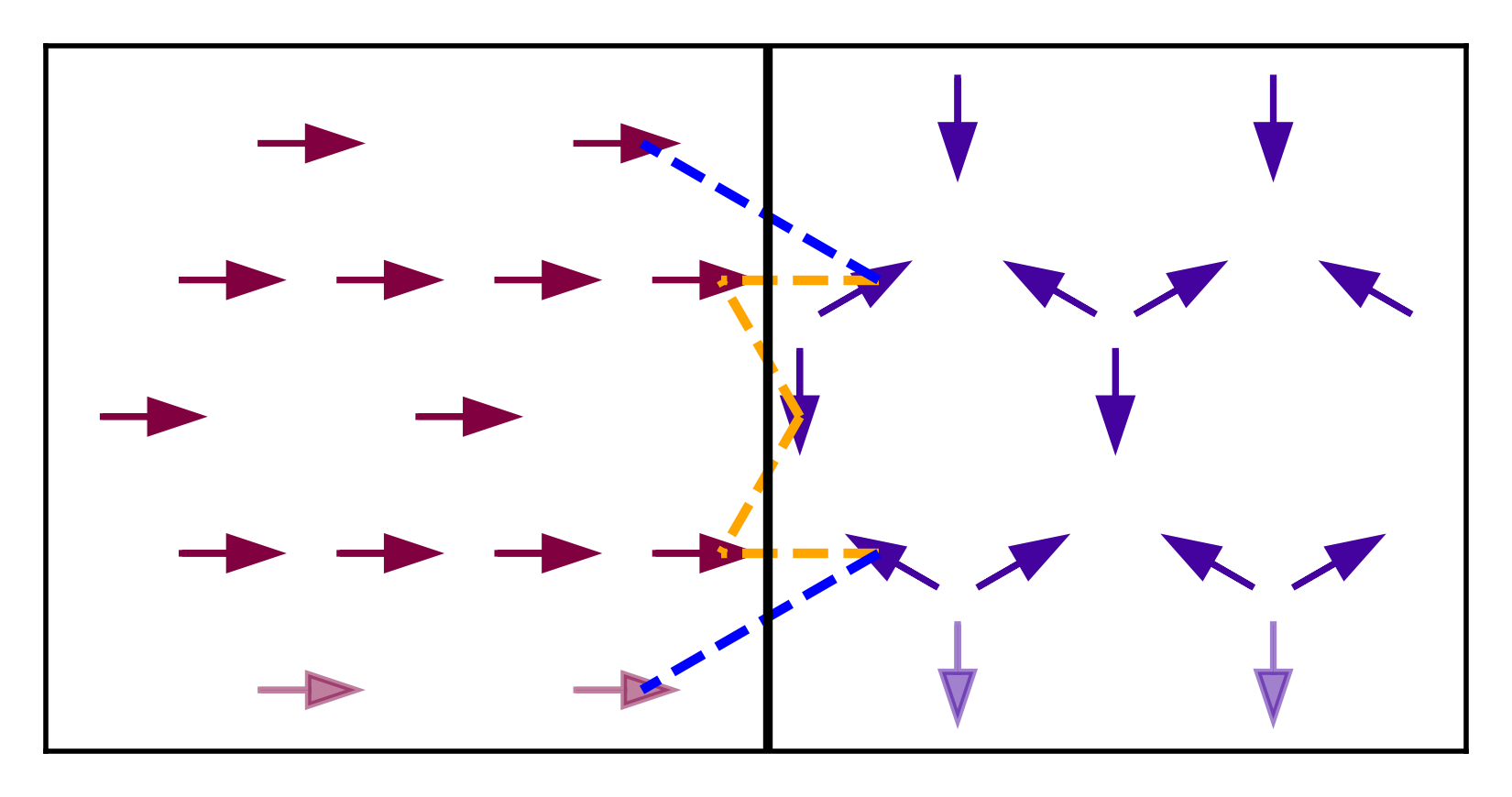}
	\caption{Schematic of the FM/AFM interface with the nearest (orange dashed lines) and next-nearest neighbor (blue dashed lines) exchange couplings across the interface. Moments at the bottom with lower opacity represent sites repeated via periodic boundary conditions.}
	\label{fig:Interface}
\end{figure}

To create a fully compensated interface in our numerical simulation, we selectively add ferromagnetic next-nearest neighbor interactions $J_{\rm nnn} = J_{\rm FM}$ to two AFM sublattices along the boundary, as illustrated in Fig.~\ref{fig:Interface}, so that the net exchange coupling between FM and AFM sites across the boundary vanishes. The above spin structure, without further relaxation, is then used as the equilibrium state for solving the linearized LLG equation. Despite being a bit unrealistic, such a minimal, sharp interface model facilitates the demonstration of the optimal boundary condition mentioned above in a controlled manner (see below). At a real FM/AFM interface, local materials details, such as lattice mismatch, magnetic structure distortion or texture, interface anisotropy, crystal orientation, atomic inter-mixing, etc. can influence the size and polarization of transmitted spin waves to different extents. But such effects can all, in principle, be described as non-universal effective interface interaction terms in a scattering formulation \cite{supp}. Therefore, the general criteria pointed out above should qualitatively hold even for real interfaces. 

\begin{figure}[ht]
	\centering
	\subfloat[]{\includegraphics[width=0.35\linewidth]{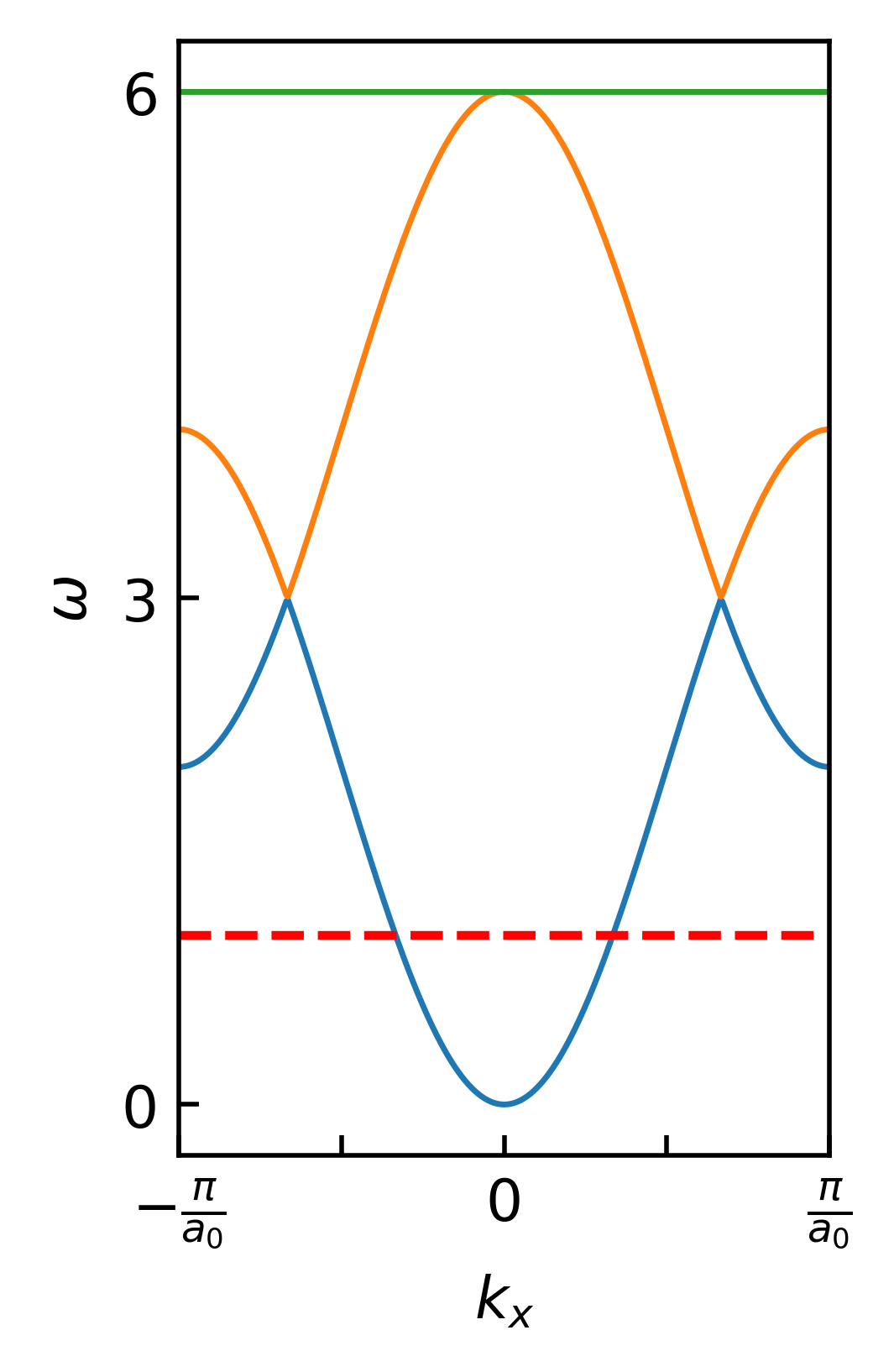}}
	\subfloat[]{\includegraphics[width=0.32\linewidth]{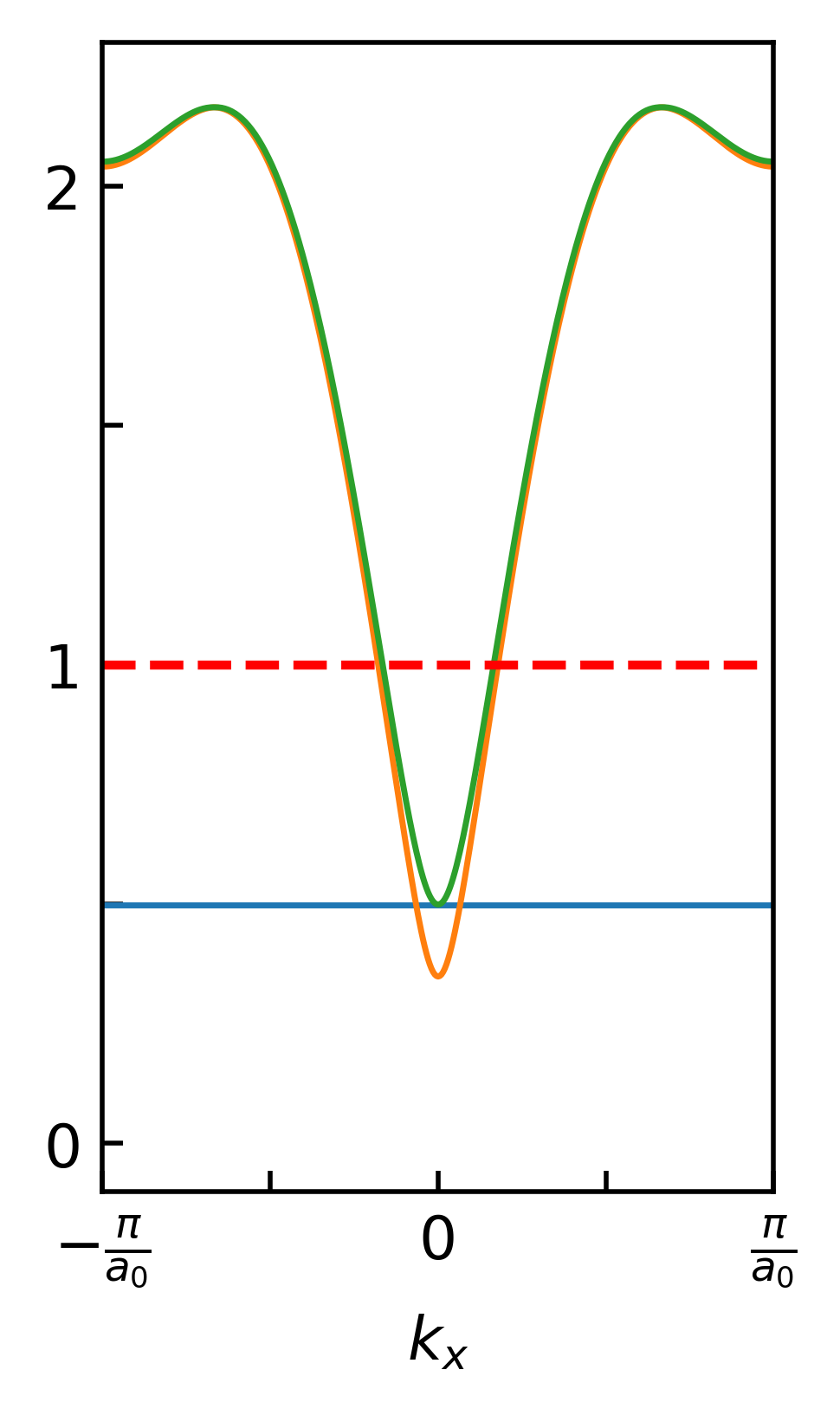}}
	\subfloat[]{\includegraphics[width=0.314\linewidth]{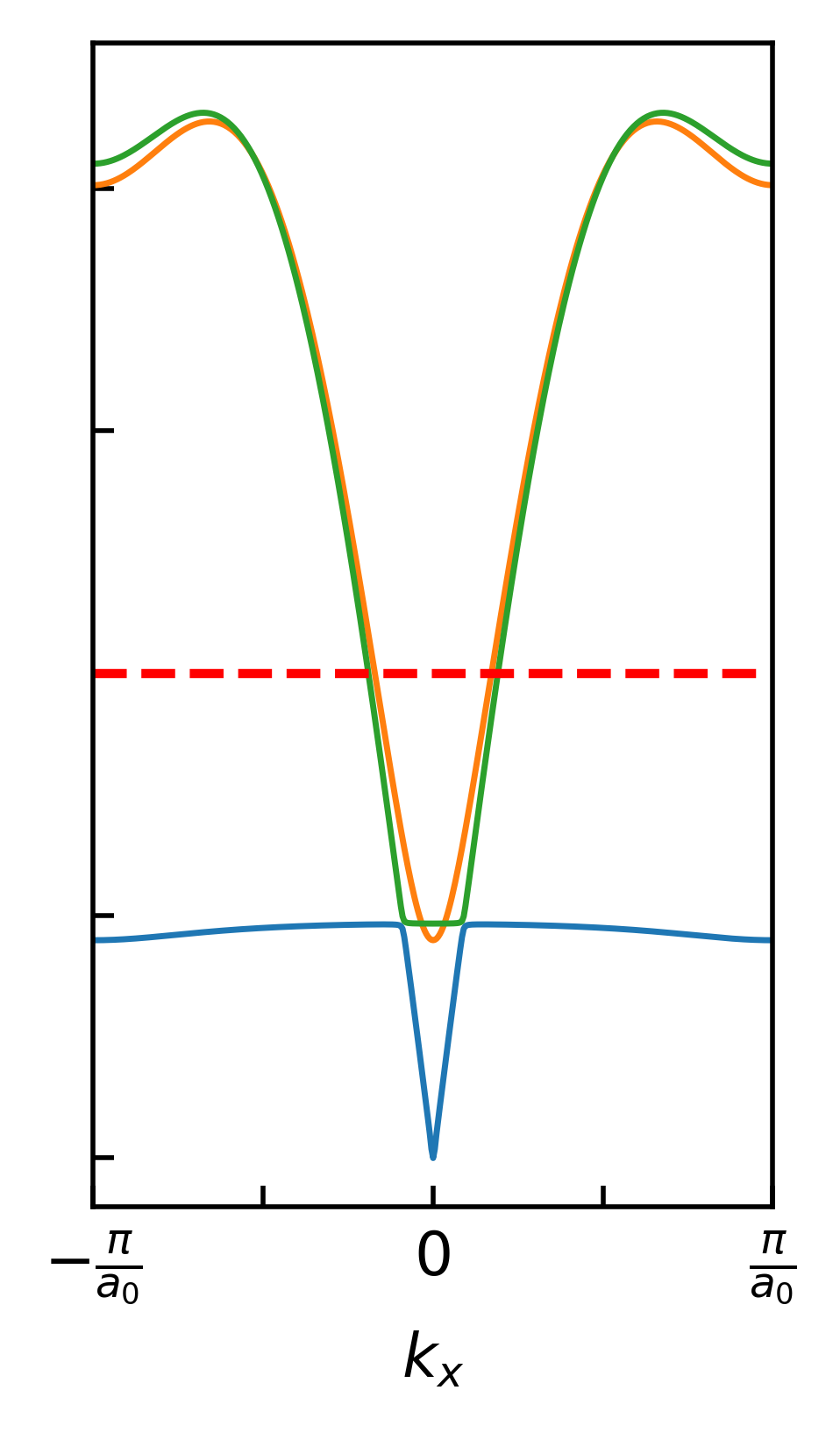}}
	\caption{Spin wave dispersions of (a) the Kagome FM, (b) and (c) the kagome AFM with direct and inverse triangular orders. Dotted red lines denote the external field frequency and targeted spin wave modes.}
	\label{Fig:Spinwavedisp}
\end{figure}

In addition, the 1D scattering picture suggests that the frequency of the a.c. driving field should be chosen so that at this frequency the desired doubly-degenerate, linearly dispersive spin wave modes that can carry spin angular momentum in the noncollinear AFM exist. To this end, we plot in Fig.~\ref{Fig:Spinwavedisp} the bulk spin waves in the FM and the two noncollinear AFM along $x$ direction, with the desired frequency $\omega = 1$ determined by the above criteria indicated by the dashed lines. 

The linearized LLG equation is then solved for our strip model with the following external a.c. magnetic field localized near the left end
\begin{eqnarray}
	\mathbf{h}^{\rm ext}_i (t)&=& \hat{y} e^{-\frac{(\mathbf{r}_i \cdot \hat{x})^2}{2\sigma^2}} e^{-i\omega t}
\end{eqnarray}
where the Gaussian width $\sigma = 2$. To avoid reflections of spin waves at the two open boundaries of our 1D system, a large damping, $\alpha = 100$ is used for spins near the boundaries ($\alpha = 0.02$ in the bulk). Note that this is necessary--without such sinks on the boundaries, spin currents will not flow. $\gamma = 1$ in all calculations. The numerical solutions of the spin waves allow us to calculate the d.c. dynamical spin currents and the driving force in the strip according to explicit lattice formulas derived from Eq.~\eqref{Eqn:DynJ} in Sec. IV of \cite{supp}. 

When plotting the relevant quantities in Fig.~\ref{Fig:Iruni}, we have grayed out an ``interface region" that is 10 unit cells wide (5 unit cells on either side) at the interface, since the spin waves have strong resonances immediately at the boundary which obscure the smooth profiles of the plotted quantities inside the bulk of the two layers. 

\begin{figure}[h]
	\centering
	\includegraphics[width=0.65\linewidth]{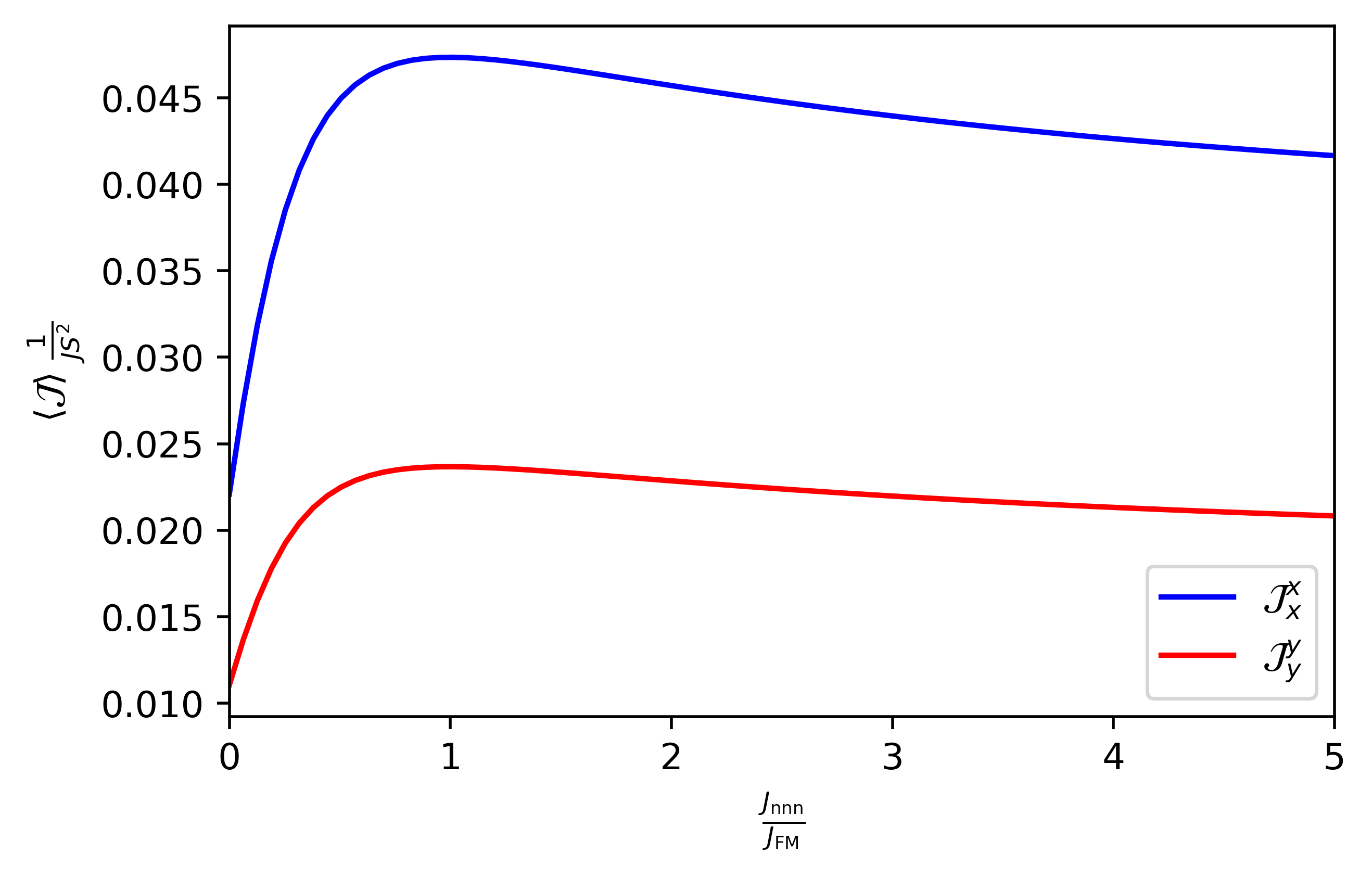}
	\caption{Spin currents at a site on the AFM side versus the next-nearest neighbor exchange coupling $J_{\rm nnn}$ scaled to the nearest neighbor value $J_{\rm FM}$. $J_{\rm nnn}/J_{\rm FM} =1$ corresponds to a fully compensated AFM interface.}
	\label{fig:nnncoupling}
\end{figure}

Finally, to see that a fully compensated AFM surface is indeed ideal for spin current transmission, we plot in Fig.~\ref{fig:nnncoupling} two major components of the spin current at a site on the AFM side versus the next-nearest neighbor exchange coupling $J_{\rm nnn}$ for the case of direct triangular order. The spin current reaches a maximum when $J_{\rm nnn} = J_{\rm FM}$ as expected.  

\bibliography{NoncollAFM_ref}

\end{document}